\documentclass[conference]{IEEEtran}
\IEEEoverridecommandlockouts
\usepackage{cite}
\usepackage{amsmath,amssymb,amsfonts}
\usepackage{graphicx}
\usepackage{textcomp}
\usepackage{xcolor}
\usepackage{gensymb}
\usepackage{algorithm}
\usepackage[noend]{algpseudocode}
\usepackage{multirow}
\usepackage{algorithm,algpseudocode}
\usepackage[margin=0.7in]{geometry}
\usepackage{mathtools}

\renewcommand{\baselinestretch}{0.94}
\usepackage[font=small,skip=2pt]{caption}

\begin{document}

\font\myfont=cmr15 at 19pt
\title{{\myfont TCD-NPE: A Re-configurable and Efficient Neural Processing Engine, Powered by Novel Temporal-Carry-deferring MACs}}

\author{Ali Mirzaeian, Houman Homayoun, Avesta Sasan \\  George Mason University, Fairfax, VA, USA \\ amirzaei@gmu.edu, hhomayoun@ucdavis.edu, asasan@gmu.edu}

\maketitle

\begin{abstract}
	In this paper, we first propose the design of Temporal-Carry-deferring MAC (TCD-MAC) and illustrate how our proposed solution can gain significant energy and performance benefit when utilized to process a stream of input data.  We then propose using the TCD-MAC to build a reconfigurable, high speed, and low power Neural Processing Engine (TCD-NPE).  We, further, propose a novel scheduler that lists the sequence of needed processing events to process an MLP model in the least number of computational rounds in our proposed TCD-NPE. We illustrate that our proposed TCD-NPE significantly outperform similar neural processing solutions that use conventional MACs in terms of both energy consumption and execution time. \vspace{-1mm}
\end{abstract}

\section{Introduction and Background} \label{intro}

Deep neural networks (DNNs) has attracted a lot of attention over the past few years, and researchers have made tremendous progress in developing deeper and more accurate models for a wide range of learning-related applications \cite{ alexnet, vgg}. The desire to bring these complex models to resource-constrained hardware platforms such as Embedded, Mobile and IoT devices has motivated many researchers to investigate various means of improving the DNN models' complexity and computing platform's efficiency \cite{sze2017efficient}.  In terms of model efficiency, researchers have explored different techniques including quantization of weights and features \cite{courbariaux2015binaryconnect, han2015deep}, formulating compressed and compact model architectures \cite{howard2017mobilenets, icnn,icnntecs,  iandola2016squeezenet, han2015deep, 8697497 }, increasing model sparsity and pruning \cite{yang2017designing, han2015deep}, binarization \cite{hubara2016binarized, courbariaux2015binaryconnect}, and other model-centered alternatives.   

On the platform (hardware) side, the GPU solutions have rapidly evolved over the past decade and are considered as a prominent mean of training and executing DNN models. Although GPU has been a real energizer for this research domain, its is not an ideal solution for efficient learning, and it is shown that development and deployment of hardware solutions dedicated to processing the learning models can significantly outperform GPU solution. This has lead to the development of Tensor Processing Units (TPU)  \cite{abadi2016tensorflow}, Field Programmable Gate Array (FPGA) accelerator solutions \cite{lacey2016deep}, and many variants of dedicated ASIC solutions \cite{diannao, dadiannao, du2015shidiannao, eyeris}.

Today, there exist many different flavors of ASIC neural processing engines. The common theme between these architectures is the usage of a large number of simple Processing Elements (PEs) to exploit the inherent parallelism in DNN models. Compare to a regular CPU with a capable Arithmetical Logic Unit (ALU), the PE of these dedicated ASIC solutions is stripped down to a simple Multiplication and Accumulation (MAC) unit. However, many PEs are used to either form a specialized data flow \cite{dadiannao}, or tiled into a configurable NoC for parallel processing DNNs \cite{eyeris}. The observable trend in the evolution of these solutions starting from DianNao \cite{diannao}, to DaDianNao \cite{dadiannao}, to ShiDianNao \cite{du2015shidiannao}, to Eyris \cite{eyeris} (to name a few) is the optimization of data flow to increase the re-use of information read from memory, and to reduce the data movement (in NOC and to/from memory).  

Common between previously named ASIC solutions, is designing for data reuse in NOC level but ignoring the possible optimization of the PE's MAC unit. A conventional MAC operates on two input values at a time, computes the multiplicaiton result, adds it to its previously accumulated sum and output a new and \textit{correct} accumulated sum. When working with streams of input data, this process takes place for every input pair taken from stream. But in many applications, we are not interested in the correct value of intermediate partial sums, and we are only interested in the correct final result. 
The first design question that we answer in this paper is if we can design a faster and more efficient MAC, if we remove the requirement of generating a correct intermediate sum, when working on a stream of input data.

In this paper, we propose the design of Temporally-deferring-Carry MAC (TCD-MAC), and use the TCD-MAC to build a reconfigurable, high speed, and low power MLP Neural Processing Engine (NPE). We illustrated that TCD-MAC can produce an approximate-yet-correctable result for intermediate operations, and could correct the output in the last state of stream operation to generate the correct output. We then build a Re-configurable and specialized MLP Processing Engine using a farm of TCD-MACs (used as PEs) supported by a reconfigurable global buffer (memory) and illustrate its superior performance and lower energy consumption when compared with the state of the art ASIC NPU solutions. To remove the data flow dependency from the picture, we used our proposed  NPE to process various Fully Connected Multi-Layer Perceptrons (MLP) to simplify and reduce the number of data flow possibilities and to focus our attention on the impact of PE in the efficiency of the resulting accelerator. 


\section{RELATED WORK}\label{related_work}
The work in \cite{eyeris}, categorizes the possible data flows into four major categories: 1) No Local Reuse (NLR) where neither the PE (MAC) output nor filter weight is stored in the PE. Examples of accelerator solutions using NLR data flow include \cite{diannao, dadiannao, zhang2015optimizing}. 2) Output Stationary (OS) where the filter and weight values are input in each cycle, but the MAC output is locally stored. Examples of accelerator solutions using OS data flow include \cite{gupta2015deep, du2015shidiannao, peemen2013memory, Nesta1}. 3) Weight Stationery (WS) where the filter values are locally stored, but the MAC result is passed on. Examples of accelerators using WS data flow include \cite{sankaradas2009massively, chakradhar2010dynamically,gokhale2014240 }, and 4) Row Stationary (RS and its variant RS+) where some of the reusable MAC outputs and filter weights remain within a local group of PE to reduce data movement for computing the next round of computation. An example of accelerator using RS is \cite{eyeris}.

The OS and NLR are generic data flow and could be applied to any DNN, while the WS and RS only apply to Convolutional Neural Networks (CNN) to promote the reuse of filter weights. Hence, the type of applicable data reuse (output and/or weight) depends on the model being processed. The Multi-Layer Perceptrons (MLP) is a sub-class of NNs that has extensively used for modeling complex and hard to develop functions \cite{block1962perceptron}. An MLP has a feed-forward structure, and is comprised of three types of layers: (1) An input layer for feeding the information to the model, 2) one or more hidden layer(s) for extracting features, and (3) an output layer that produces the desired output which could be regression, classification, function estimation, etc. Unfortunately, when it comes to MLPs, or when processing Fully Connected (FC) layers, unlike CNNS, no filter weight could be reused. In these models the viable data flows are the OS and NLR. The only possible solution for using the WS solution in processing MLPs is the case of multi-batch processing that may benefit from weight reuse. 
Another related work is the NPE proposed in \cite{tu2015rna}. This solution, denoted as RNA, is a special case of NLR, where data flow is controlled through NoC connectivity between different PEs; 
RNA breaks the MLP model into multi-layer loops that are successively mapped to the accelerator PEs, and uses the PEs as either a multiplier or an adder, dynamically forming a systolic array.  

In the result section of this paper, We demonstrate that the OS solutions are in general more efficient than NLR solutions. We further illustrate that our proposed TCD-MAC, when used in the context of our proposed NPE, outperform state of the art accelerators that rely on (fastest and most efficient) conventional MAC solutions.

\section{Our Proposed MLP Processing Engine}

Before describing our proposed NPE solution, we first describe the concept of \emph{temporal carry} and illustrate how this concept can be utilized to build a Temporal Carry deferring Multiplication and Accumulation (TCD-MAC) unit. Then, we describe, how an array of TCD-MAC are used to design a re-configurable and high-speed MLP processing engine, and how the sequence of operations in such NPE is scheduled to compute multiple batches of MLP models.   

\subsection{Temporal Carry deferring MAC (TCD-MAC)}\label{TCD_MAC_section}

\begin{figure*}[hbt!]
	\centering
	\includegraphics[width=1.80\columnwidth]{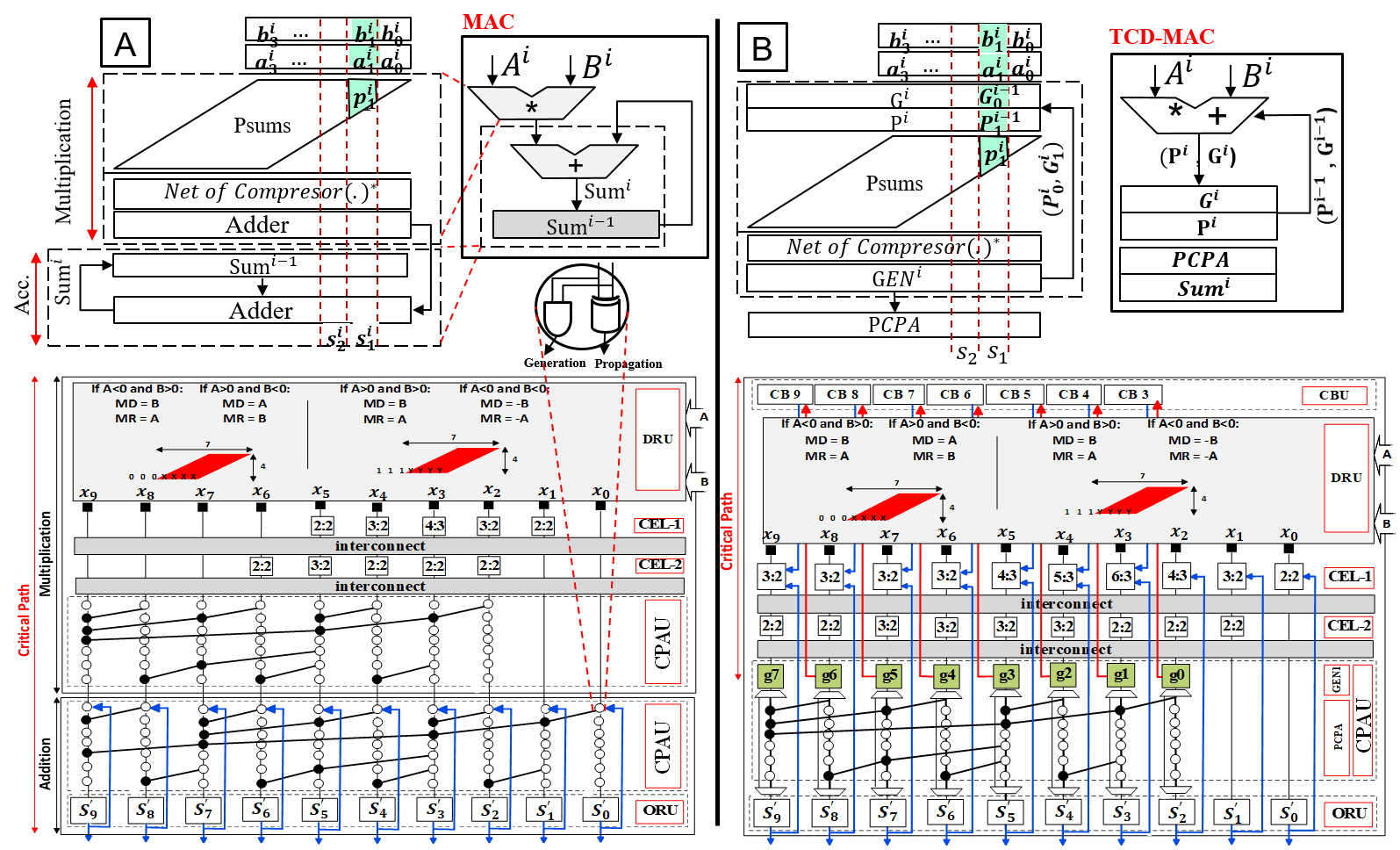}
	\caption{Comparing the architecture of A) a typical MAC, versus B) a simplified 2-input version of TCD-MAC. In all variables in form of $D^i_m$, the subscript ($m$) captures the bit position values, and postscript ($i$) capture the cycle (iteration). For example, $A^i, B^i$ are the input data 
		in the $i^{th}$ iteration (corresponding to the $i^{th}$ cycle) of the multiply accumulate operation. The $b_{m}^{i}, a_{m}^{i}$, and $p_{m}^{i}$ are accordingly the $m^{th}$ significant bits of inputs $A$, $B$, and partial sum at the $i^{th}$ cycle (iteration). The division of CPA into GEN and PCPA is also shown in this figure. Note that the $PCPA$ is only executed at the last cycle.\vspace{-5mm}}
	\label{GV_TCD-MAC}
\end{figure*}

Suppose two vectors $A$ and $B$ each have $N$ M-bit values, and the goal is to compute their dot product, $\sum_{i=0}^{N-1}(A_i*B_i)$ (similar to what is done during the activation process of each neuron in a NN). This could be achieved using a single Multiply-Accumulate (MAC) unit, by working on 2 inputs at a time for N rounds. Fig. \ref{GV_TCD-MAC}(A-top) shows the general view of a typical MAC architecture that is comprised of a multiplier and an adder (with 4-bit input width), while Fig. \ref{GV_TCD-MAC}(A-bottom) provides a more detailed view of this architecture. The partial products (M partial product for M-bits) are first generated in Data Reshape Unit (DRU). Then the hamming weight compressors (HWC) in the Compression and Expansion Layer (CEL) transform the addition of M partial products into a single addition of two larger binaries, the addition of which in an adder generates the multiplication result.

The building block of the CEL unit are the HWC. A HWC, denoted by C$_{HW}$(m:n), is a combinational logic that implements the Hamming Weight (HW) function for $m$ input-bits (of the same bit-significance value) and generates an $n$-bit binary output. The output $n$ of HWC is related to its input $m$ by: $n = \lceil log_{2}^{m}\rceil$. For example "011010", "111000", and "000111" could be the input to a C$_{HW}$(6:3), and all three inputs generate the same Hamming weight value represented by "011". A Completed HWC function CC$_{HW}$(m:n) is defined as a C$_{HW}$ function, in which  $m$ is $2^{n}-1$ (e.g.,  CC(3:2) or CC(7:3)). Each HWC takes a column of m input bits (of the same significance value) and generates its n-bit hamming weight. In the CEL unit, the output n-bits of each HWC is fed (according to its bit significance values) as an input to the proper $C_{HW}$(s) in the next-layer CEL. This process is repeated until each column contains no more than 2-bits, which is a proper input size for a simple adder.  In Fig. \ref{GV_TCD-MAC} it is assumed that a Carry Propagation Adder Unit (CPAU) is used. The result is then added to the previously accumulated value in the output register in the second adder to generate a new accumulated sum. Note that in conventional MAC, the carry (propagation) bits in the CPAUs are spatially propagated through the carry chain which constitutes the critical timing path for both adder and multiplier.

Fig.\ref{GV_TCD-MAC}.B shows our proposed TCD-MAC.  In this solution, only a single CPAU is used. Furthermore, the CPAU is broken into two distinct segments 1) The GENeration (GEN) and Partial CPA (PCPA). The Gen is the first layer of CPA logic that produces the Generate ($G_i^c$) and Propagate ($P_i^c)$ signals for each bit position $i$ at cycle $c$. The TCD-MAC relies on the assumption that we only need to correctly compute the final result of multiplication and accumulation over an array of inputs (e.g. $\sum_{i=0}^{N-1}(A_i*B_i)$), while relaxing the requirement for generating correct intermediate sums. This relaxed specification is applicable when a MAC is used to compute a Neuron value in a DNN. Benefiting from this relaxed requirement, the TCD-MAC skips the computation of PCPA, and injects (defers) the $G_i^c$ and $P_i^c$ generated in cycle c, to the CEL unit in cycle $c+1$. Using this approach, the propagation of carry-bit in the long carry chain (in PCPA) is skipped, and without loss of accuracy, the impact of the carry bit is injected to the correct bit position in the next cycle of computation. We refer to this process as temporal (in time) carry propagation. The Temporally carried $G_i^c$ is stored in a new set of registers denoted as Carry Buffer Unit (CBU), while the $P_i^c$ in each cycle is stored in the output register Unit (ORU). Note that CBU bits can be injected to any of the $C_{HW}(m:n)$ in any of the CEL layers in the same bit position. However, it is desired to inject the CB bits to a $C_{HW}(m:n)$ that is incomplete to avoid an increase in the size and critical path delay of the CEL. 
\begin{figure}[h]
	\centering
	\includegraphics[width=0.72\columnwidth]{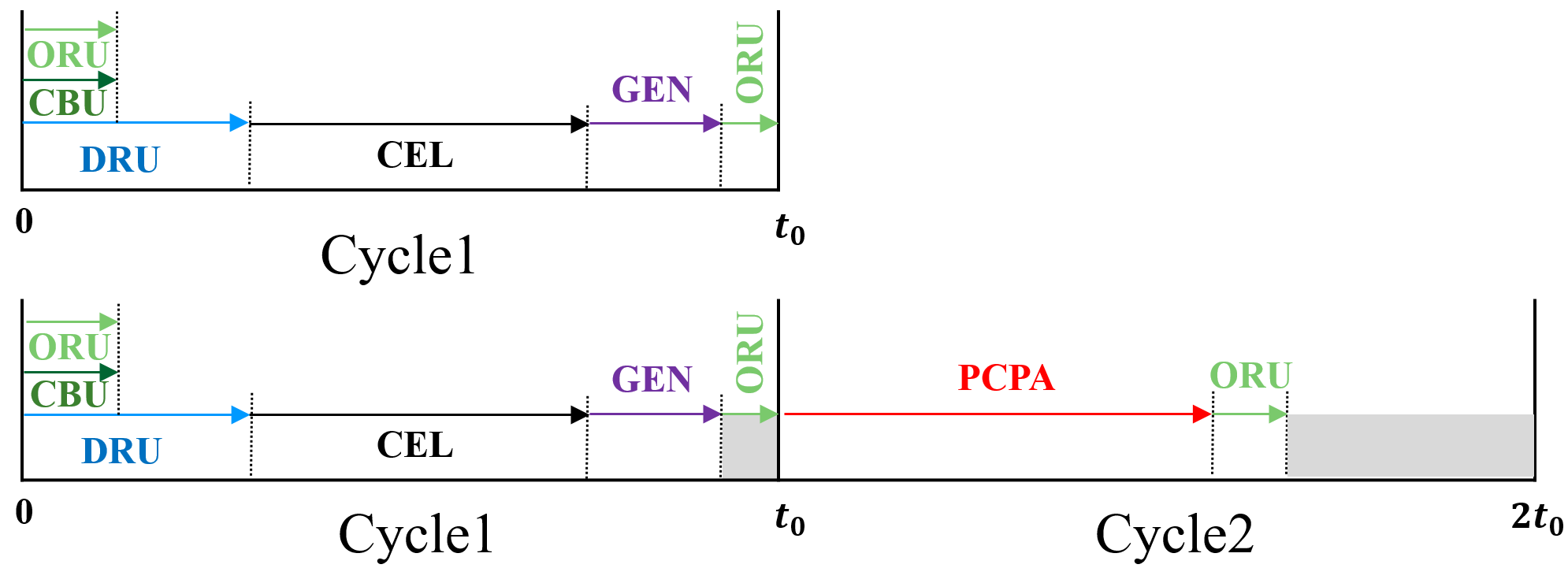}
	\caption{TCD-MAC cycle time is computed by excluding the PCPA. In the last cycle of computation, the TCD-MAC activates the PCPA to propagate the unconsumed carry bits.\vspace{-6mm} }
	\label{cycle_reduction}
\end{figure}


Assuming that a TCD-MAC works on an array of N input pairs, the temporal carry injection is done N-1 times. In the last round, however, the PCPA should be executed. As illustrated in Fig. \ref{cycle_reduction}, in this approach, the cycle time of the TCD-MAC could be reduced to that excluding the PCPA, allowing the computation over PCPA to take place in an extra cycle. The one extra cycle allows the unconsumed carry bits to be propagated in PCPA carry chain, forcing the TCD-MAC to generate the correct output. Using this technique we shortened the cycle time of TCD-MAC for a large number of cycles. The saving obtained from shorter cycles over a large number of cycles significantly outweighs the penalty of one extra cycle.


To support signed inputs, in TCD-MAC we pre-process the input data. For a partial product $p=a\times b$, if one value ($a$ or $b$) is negative, it is used as the multiplier. With this arrangement, we treat the generated partial sums as positive values and later correct this assumption by adding the two's complement of the multiplicand during the last step of generating the partial sum. Following example clarify this concept: let's suppose that $a$ is a positive and $b$ is a negative b-bit binary. The multiplication $b\times a$ can be reformulated as: 

\vspace{-3 mm}
\begin{equation}
	\scriptsize
	b \times a = (-2^{7}+\sum_{i=0}^{6}x_{i}2^{i}) \times a= -2^{7}a+(\sum_{i=0}^{6}x_{i}2^{i}) \times a
	\vspace{-2mm}
\end{equation}

The term $-2^{7}a$ is the two's complement of multiplicand which is lef-shifted by 7 bits, and the term ($\sum_{i=0}^{6}x_{i}2^{i}) \times a$ is only accumulating shifted version of the multiplicand.

\subsection{TCD-NPE: Our Proposed MLP Neural Processing Engine\vspace{-1mm}}\label{TCD_NPE_section}

TCD-NPE is a configurable neural processing engine which is composed of a 2-D array of TCD-MACs. The TCD-MAC array is connected to a global buffer using a configurable Network on Chip (NOC) that supports various forms of data flow as described in section \ref{intro}. However, for simplicity, we limit our discussion to supporting OS and NLR data flows for executing MLPs. This choice is made to help us focus on the performance and energy impact of utilizing TCD-MACs in designing an efficient NPE without complicating the discussion with the support of many different data flows. 

Figure \ref{npe_architecture} captures the overall TCD-NPE architecture. It is composed of 1) Processing Element (PE) array which is a tiled array of TCD-MACs, 2) Local Distribution Networks (LDN) that manages the PE-array connectivity to memories, 3) Two global buffers, one for storing the filter weights and one for storing the feature maps, and 4) The Mapper-and-controller unit which translates the MLP model into a supported data and control flow. The functionality and design of each of these units are described next:

\begin{figure}[!hbt]
	\centering
	\includegraphics[width=0.9\columnwidth]{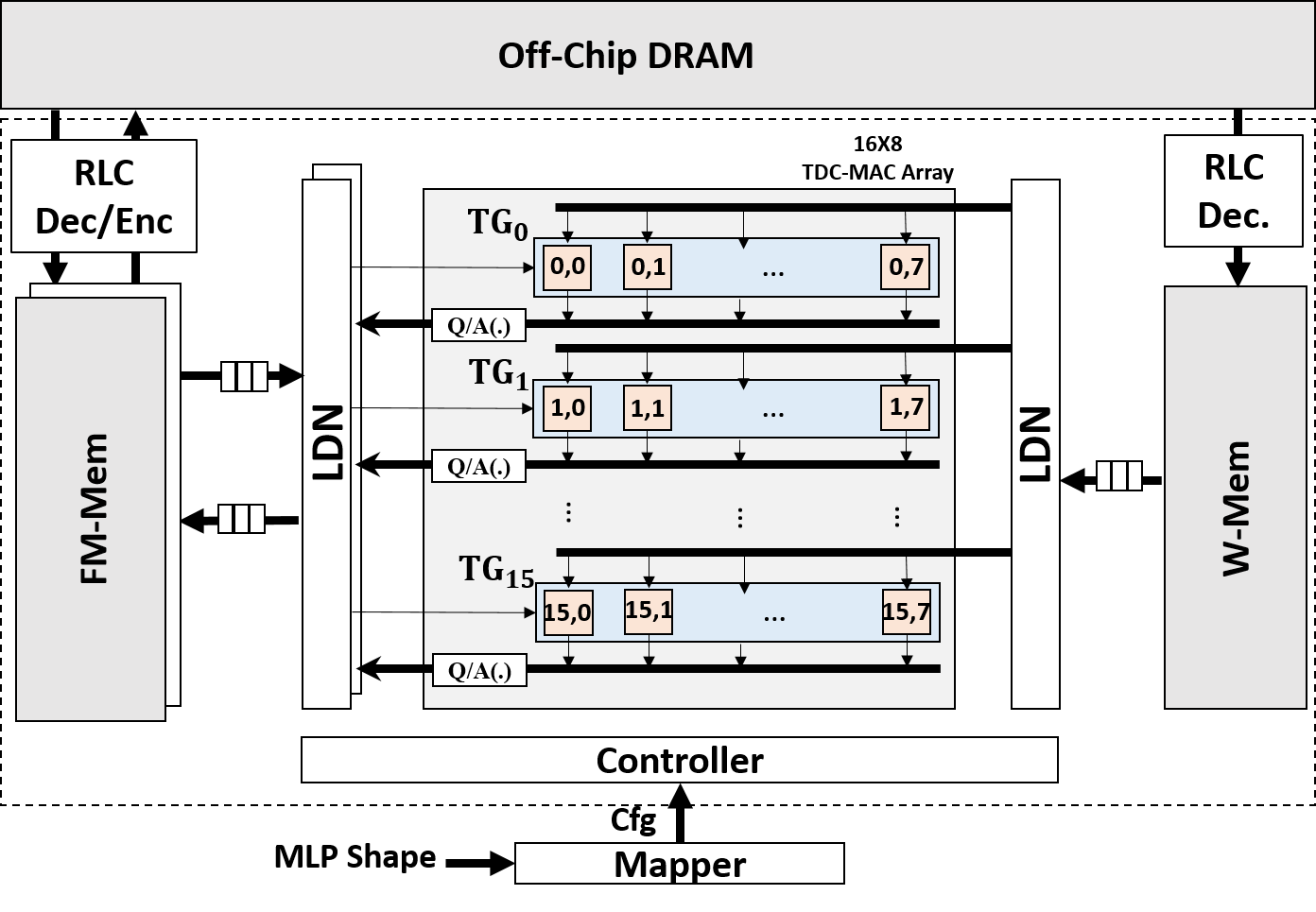}
	\caption{TCD-NPE overall architecture. The Mapper algorithm is executed externally, and the sequence of events is loaded into the controller for governing the OS data and control flow.\vspace{-4mm}}
	\label{npe_architecture}
\end{figure}

\subsubsection{\textbf{PE Array}}
\begin{figure}[t]
	\centering
	\includegraphics[width=0.80\columnwidth]{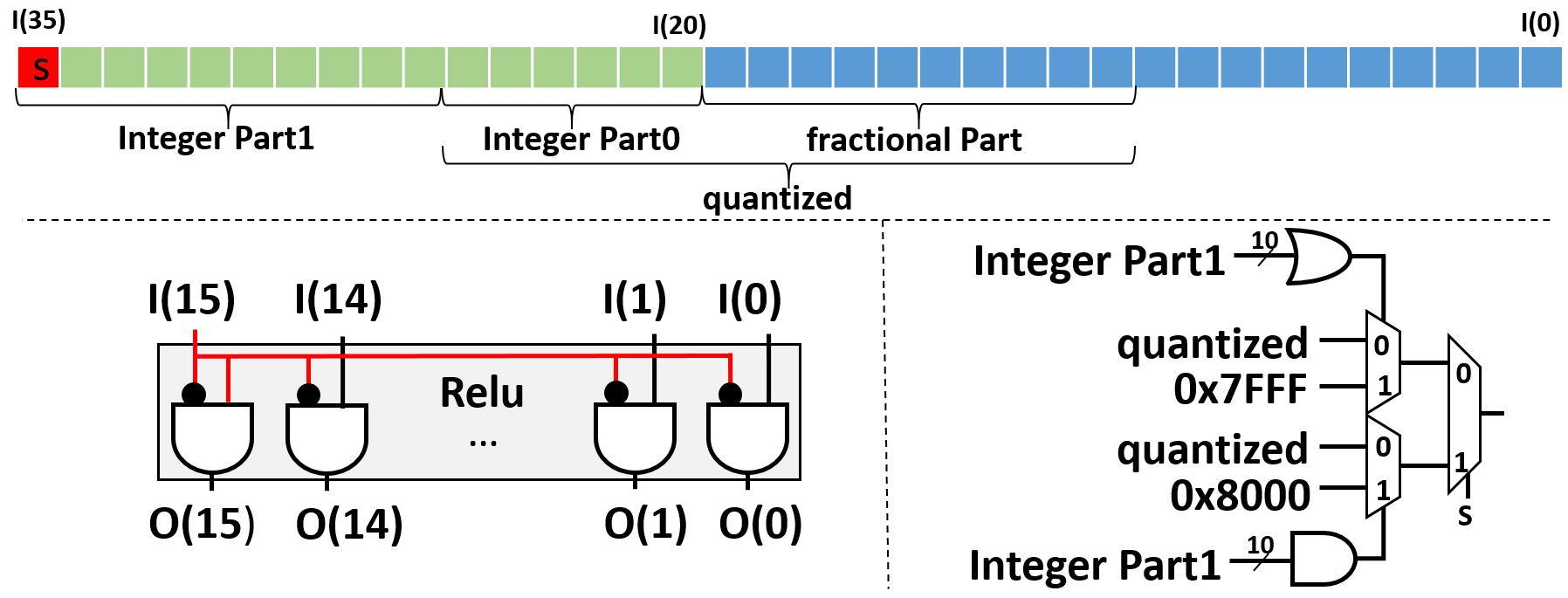}
	\caption{The logic implementation of Quantization (Left) and Relu Activation (right) for signed fixed-point 16bit values \vspace{-4mm}}
	\label{sequence}
	\vspace{-2 mm}
\end{figure}

The PE-array is the computational engine of our proposed TCD-NPE. Each PE in this tiled array is a TCD-MAC. Each TCD-MAC could be operated in two modes: 1) Carry Deferring Mode (CDM), or 2) Carry Propagation Mode (CPM). According to the discussion in section \ref{TCD_MAC_section}, when working with an input stream of size N, the TCD-MAC is operated in the CDM model for N cycles (computing approximate sum), and in the CPM mode in the last cycle to generate the correct output. This is in line with OS data flow as described in section \ref{related_work}. Note that the TCD-MAC in this PE-array could be operated in CPM mode in every cycle allowing the same PE-array architecture to also support the NLR. After computing the raw neuron value (prior to activation), the TCD-MAC writes the computed sum into the NOC bus. The Neuron value is then passed to the quantization and activation unit before being written back to the global buffer. Fig.  \ref{sequence} captures the logic implementation for quantization (to 16 bits) and Relu\cite{alexnet} activation in this unit. 

Consider two layers of an MLP where the input layer contains M feature-values (neurons) and the second layer contains N Neurons. To compute the value of N Neurons, we need to utilize N TCD-MACs (each for M+1 cycles). If the number of available TCD-MACS is smaller than N, the computation of the neurons in the second layer should be unrolled to multiple rolls (rounds). If the number of available TCD-MACs is larger than neurons in the second layer (for small models), we can simultaneously process multiple batches (of the model) to increase the NPE utilization. Note that the size of the input layer (M) will not affect the number of needed TCD-MACs, but dictates how many cycles (M+1) are needed for the computation of each neuron. 


When mapping a batch of MLP to the PE-array, we should decide how the computation is unrolled and how many batches (K), and how many output neurons (N) should be mapped to the PE-array in each roll. The optimal choice would result in the least number of rolls and the maximum utilization of the NPE. To illustrate the trade-offs in choosing the value of (K, N) let us consider a PE-array of size 18, which is arranged in 6 rows and 3 columns of TCD-MACs (similar to that in Fig. \ref{npe_architecture}). We refer to each row of TCD-MACs as a TCD-MAC Group (TG). In our implementation, to reduce NOC complexity, the TG groups work on computing neurons in the same batch, while different TG groups could be assigned to work on the same or different batches. The architecture in Fig. \ref{npe_architecture} has 6 TG groups. Let us use NPE(K, N) to denote the choice of using the PE-array to compute N neuron values in K batches where $N\times K=18$. In our example PE-array the following selections of K and N are supported: $(K,N) \in (1,18), (2, 9), (3, 6), (6, 3)$. The $(9,2)$ and $(18,1)$ configuration are not supported as the value of N in this configurations is smaller than TG size = 3. 

\begin{figure}[h]
	\centering
	\includegraphics[width=0.85\columnwidth]{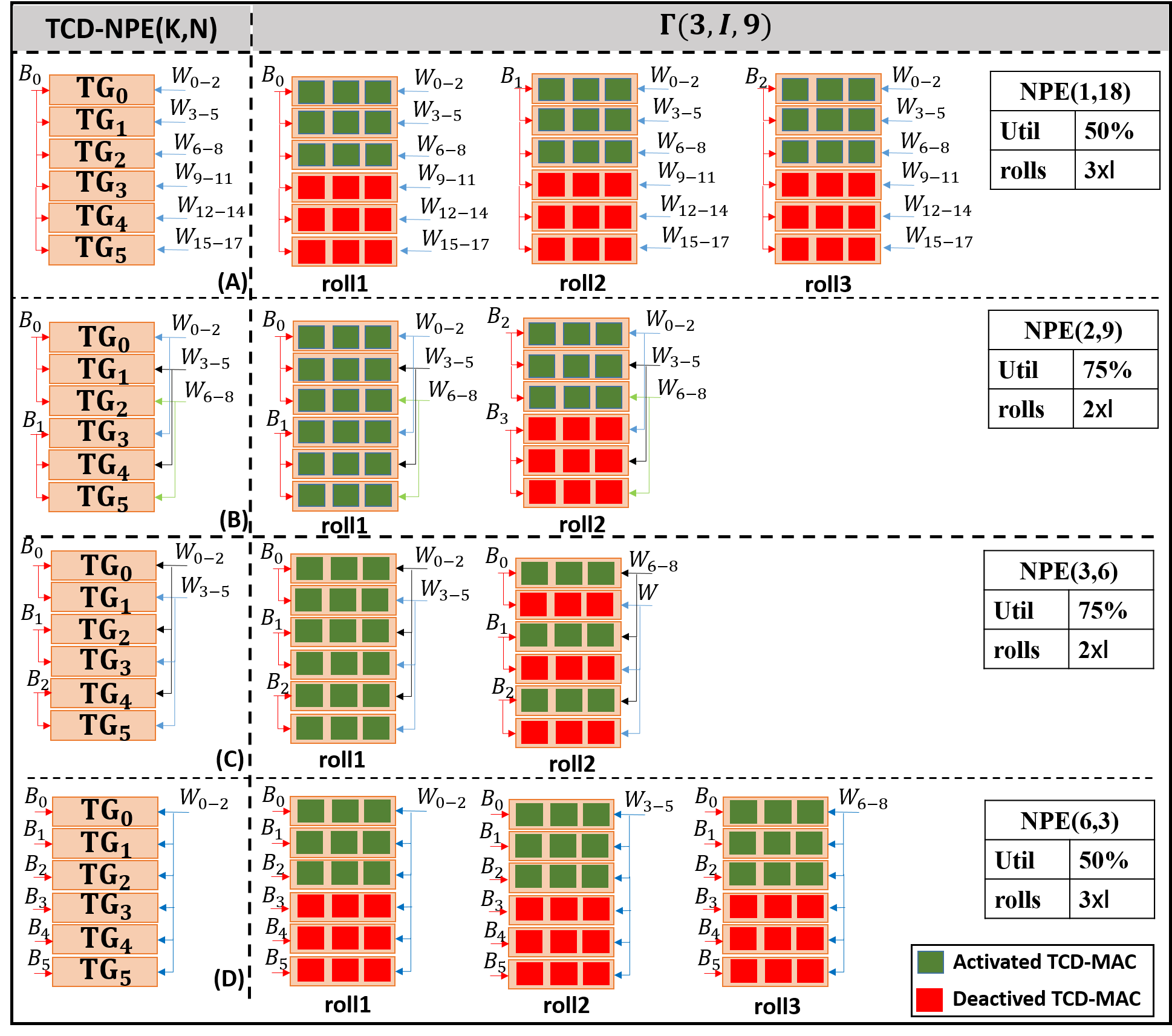}
	\caption{Assuming a $6 \times 3$ PE-array of TCD-MACs, the NPE(K, N) could be configured such that (K, N) $\in$ \{(1,18), (2,9), (3,6), (6,3)\}. This figure illustrate the number of rolls, and utilization when each of NPE(K,N) configurations is used to run a $\Gamma$(3,I,9). model. Each roll is executed I times.\vspace{-2mm}}
	\label{TCD_NPE}
\end{figure}

Fig. \ref{TCD_NPE}.left shows an abstract view of TCD-NPE and describe how the weights and input features (from one or more batches) are fed to the TCD-NPE for different choices of K and N. As an example \ref{TCD_NPE}.(left).A shows that input features from one batch are broadcasted between all TGs, while the weights are unicasted to each TCD-MAC. Let us represent the input scenario of processing B batches of U neurons in a hidden or output layer of an MLP model with I input features using $\Gamma(B,I,U)$. Fig. \ref{TCD_NPE}.(right) shows the NPE status when a $\Gamma(3,I,9)$ model (3 batches of a hidden layer with 9 neurons in a hidden layer each fed from I input neurons) is executed using each of 4 different NPE(K, N) choices. For example Fig. \ref{TCD_NPE}.(right).top shows that using configuration NPE(1,18), we process one batch with 18 neurons at a time. In this example, when using this configuration, the NPE is underutilized (50\%) as there exist only 9 neurons in each batch. Following a similar argument, the NPE(6,3) arrangement also have 50\% utilization. However the arrangement NPE(2,9), and NPE(3,6) reach 75\% utilization (100\% for the roll, and 50\% for the second roll), hence either NPE(2,9) or NPE(3,6) arrangement is optimal for the $\Gamma(3,I,9)$ problem as they produce the least number of rolls. Note that the value of I in $\Gamma(3,I,9)$ denotes the number of input features which dictate the number of cycles that the NPE(K,N) should be executed.\vspace{-2MM}

\subsubsection{\textbf{Mapping Unit}}
An MLP has one or more hidden layers and could be presented using Model($I-H_1-H_2-...-H_N-O$), in which $I$ is the number of input features, $H_i$ is the number of Neurons in the hidden layer $i$, and $O$ is the number of output layer neurons.  The role of the mapping unit is to find the best unrolling scenario for mapping the sequence of problems $\Gamma(B,I, H_1)$, $\Gamma(B,H_1, H_2)$, ..., $\Gamma(B,H_{N-1}, H_N)$, and $\Gamma(B, H_N,O)$ into minimum number of NPE(K,N) computational rounds. 
 
Algorithm \ref{mapper_algorithm} describes the mapper function for unrolling a multi-batch multi-layer MLP problem. In this Algorithm, B is the batch size that could fit in the NPE's feature-memory (if larger, we can unroll the B into N $\times$ B* computation round, where B* is the number of batches that fit in the memory). $M[L]$ is the MLP layer size information, where $M[i]$ is the number of nodes in layer $i$ (with $i=0$ being Input, and $i=N+1$ being Output, and all others are hidden layers). The algorithm schedules a sequence of NPE(K, N) events to compute each MLP layer across all batches.   

\begin{algorithm}
	
	\caption{Schedule NPE(K,N) rolls (events) to execute $B$ batches of $M(L) = MLP(I, H_1,..., H_N, O)$.}
	\label{mapper_algorithm}
	\begin{algorithmic}
		\scriptsize
		
		\Procedure{PracticalCfgFinder}{Model $M[L]$, BatchSize $B$}
		\For{$(l=1; size(M); l++)$}
		\State $Tree_{head} = CreateTree(B, M[l])$
		\State $Exec_{Tree} \gets$ Shallowest binary tree (least rolls) from $Tree_{head}$  
		\State \textbf{Schedule} $\gets$ Schedule computational events by using BFS \\
		\hspace*{7em} on $Exec_{Tree}$ to report NPE(K,N) and $r$ at each node.
		\EndFor
		\State return \textbf{Schedule}
		\EndProcedure\\ 
		
		\Procedure{CreateTree}{$B, \Theta$} 
		\State $C[i] \gets$ find each $(K_i, N_i)| K_i, N_i \in \mathbb{N}$,  \& $K_i < B$  \\
		\hspace*{13em}  \& $size(NPE) = K_i\times N_i$
		\For{$(i=0; i< size(C); i++)$}
		\State $M_B = min(B, C[i][1])$. \Comment{\textcolor{blue}{C[i][1] = $K_i$}}
		\State $M_{\Theta} = min(\Theta, C[i][2])$.  \Comment{\textcolor{blue}{C[i][2] = $N_i$}}
		\State $\psi = (M_B, M_{\Theta})$   \Comment{\textcolor{blue}{$\psi$: NPE's (K,N) configuration}}
		\State $r = \lfloor B/M_B \rfloor \times \lfloor \Theta/M_{\Theta} \rfloor$ \Comment{\textcolor{blue}{$r$: \# of rolls with NPE($M_B, M_{\Theta})$}}
		\If{$(B\%M_B) \,\, != 0$} 
		\State $Node_B \gets$ CreateTree($B \% M_B, \Theta$)
		\EndIf
		\If{$(K\%M_{\Theta}) \,\, != 0$}
		\State $Node_{\Theta} \gets$ CreateTree($B-B\%M_B, \,\, K\%M_{\Theta}$)
		\EndIf
		\State \textbf{Node} $\gets$ $createNode(r, \psi, Node_B, Node_{\Theta})$
		        
		\EndFor
		\State return \textbf{Node}
		\EndProcedure
		
		\normalsize
	\end{algorithmic}

\end{algorithm}
\setlength{\textfloatsep}{2pt}

\begin{figure*}[!hbt]
	\centering
	\includegraphics[width=2.0\columnwidth]{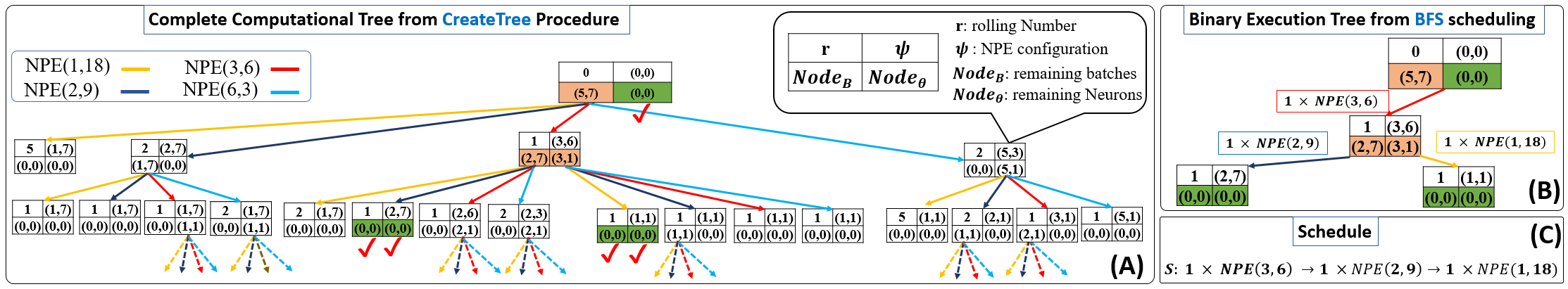}
	\caption{An example execution of algorithm \ref{mapper_algorithm} when processing $\Gamma(5,I,7)$ model using a TCD-MAC with a $6 \times 3$ PE-array. (A): the complete computational Tree from CreateTree procedure, (B): binary execution tree obtained from BFS scheduling, (C): the sequence of scheduled events to compute the model based on binary execution tree\vspace{-4mm}.}
	\label{mapping}
\end{figure*}

To schedule the sequence of events, the Alg. \ref{mapper_algorithm} first generates the expanded computational tree of the NPE using $CreateTree$ procedure. This procedure first finds all possible ways that NPE could be segmented for processing N neurons of K batches, where $K \leq B$ and stores them into configuration database C. Then for each of configurations of NPE(K, N), it derives how many rounds (r) of NPE(K, N) computations could be executed. Then it computes a) the number of remaining batches (with no computation) and b) the number of missing neurons in partially computed batches. It, then, creates a tree-node, with 4 major fields  1) the load-configuration $\Psi(K_i^*,N_i^*)$ that is used to partially compute the model using the selected NPE($K_i, N_i$) such that $(K_i^* \leq K_i) \& (N_i^* \leq N_i)$, 2) the number of rounds (rolls) $r$ taken with computational configuration $\Psi$ to reach that node, 3) a pointer to a new problem $Node_B$ that specifies the number of remaining batches (with no computation), and 4) a pointer to a new problem $Node_{\Theta}$ for partially computed batches. Then the $CreateTree$ procedure is recursively called on each of the $Node_B$ and $Node_{\Theta}$ until the batches left, and partial computation left in a (leaf) node is zero. At this point, the procedure returns. After computing the computational tree, the mapper extracts the best execution tree by finding a binary tree with the least number of rolls (where all leaf nodes have zero computation left). The number of rolls is computed by summing up the $r$ field of all computational nodes. Finally, the mapper uses a Breath First Search (BFS) on the Execution Tree ($Exec_{Tree}$ and report the sequence of $r \times $NPE(K, N) for processing the entire binary execution tree. The reported sequence is the optimal execution schedule. Fig. \ref{mapping} provides an example for executing 5 batches of a hidden MLP layer with 7 neurons. As illustrated the computation-tree (Fig. \ref{mapping}.A) is first generated, and then the optimal binary execution tree (Fig. \ref{mapping}.B) resulting in the minimum number of rolls is extracted.  Fig. \ref{mapping}.C captures the result of scheduling step where BFS search schedule the sequence of $r \times $NPE(K, N) events.

\subsubsection{\textbf{Controller}}
The controller is an FSM that receives the "Schedule" from Mapper and generated the appropriate control signals to control the proper OS data flow for executing the scheduled sequence of events. 



\begin{figure*}[h]
	\centering
	\includegraphics[width=2\columnwidth]{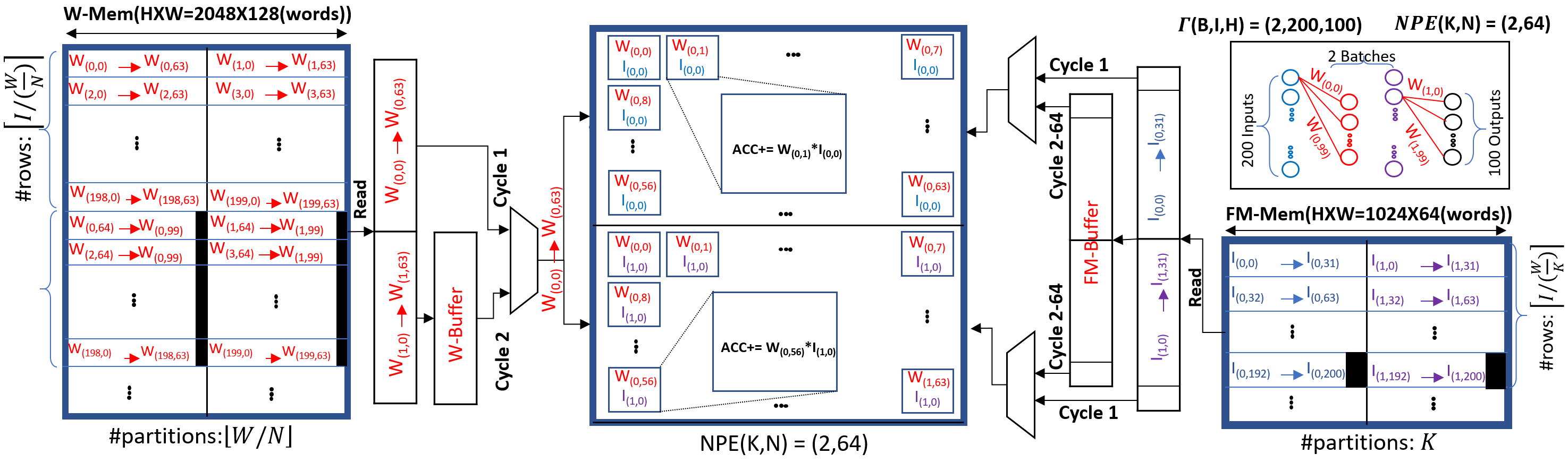}
	\caption{The arrangement of data in W-mem and FM-mem when our proposed TCD-NPE is used in NPE(K,N)=(2,64) configuration mode to process $B=2$ batches of a hidden layer of an MLP model as defined by $\Gamma(B,I,H)=(2,200,100)$. \vspace{-5mm}}
	\label{mem_access_fig}
\end{figure*}

\subsubsection{\textbf{memory architecture}}
The NPE global memory is divided into feature-map memory (FM-Mem), and Filter Weight memory (W-Mem). The FM-Mem consist of two memories with ping-pong style of access, where the input features are read from one memory, and output neurons for the next NN layer, are written to the other memory. When working with multiple batches (B), the input features from the largest number of fitting batches (B*) is read into feature memory. For simplicity, we have assumed that the feature map is large enough to hold the features (neurons) in the largest layer of at least one MLP (usually the input) layer. Note that the NPE still can be used if this assumption is violated, however, now some of the computed neuron values have to be transferred back and forth between main memory (DRAM) and the FM-Mem for lack of space. The filter memory is a single memory that is filled with the filter weights for the layer of interest.  The transfer of data from main memory (DRAM) to the W-Mem and FM-Mem is regulated using Run Length Coding (RLC) compression to reduce data transfer size and energy.

The data arrangement of features and weights inside the FM-Mem and W-Mem is shown in Fig. \ref{mem_access_fig}. The data storage philosophy is to sequentially store the data (weight and input features) needed by NPE (according to its configuration) in consecutive cycles in a single row. This data reshaping solution allows us to reduce the number of memory accesses by reading one row at a time into a buffer, and then consuming the data in the buffer in the next few cycles. We explain this data arrangement concept using the example shown in Fig. \ref{mem_access_fig}.

Fig. \ref{mem_access_fig} shows the arrangement of data when we use our proposed TCD-NPE in NPE(K,N)=(2,64) configuration to process $B=2$ batches of a hidden layer of an MLP model as defined by $\Gamma(B,I,H)=(2,200,100)$. Note that the PE array size, in this case is $16 \times 8$ which is divided into two $8 \times 8$ arrays for processing each of 2 batches. The W-Mem, shown in left, is filled by storing the first N=64 weights of each outgoing edge from input Neurons (features) to each of the neurons in the hidden layer. Considering that the width of W-Mem is 256 bytes, and each weight is 2 bytes, the width of W-Mem ($W_{W-mem}$) is 128 words. Hence, we can store 64 weights of the outgoing edge from each 2 input neurons in one row. The memory-write process is repeated for $\lceil(I/(W_{W-mem}/N))\rceil=100$ rows, and then the next $N=64$ weights of outgoing edges from each input neuron are written (in this case we only have 36 weights left, as there exist a total of 100 outgoing edges from each input neuron, 64 of which is previously stored) in the next $\lceil(I/(W_{W-mem}/N))\rceil=100$ rows. At processing time, by using the NPE(2,64) configuration, the TCD-NPE consumes $N=64$ weights in each cycle. Hence, with one read from W-Mem, it receives the weights needed for $W_{W-mem}/N=128/64=2$ cycles, reducing the number of memory accesses by half.   

The FM memory, on the other hand, is divided into $B=2$ segments. Assuming that the width of FM memory is $W_{FM-mem}=64$ words, each segment can store $W_{FM-mem}/B=64/2=32$ input features. The memory, as shown in Fig. \ref{mem_access_fig}, is filled by writing the input features of each batch into subsequent rows of each virtually segmented memory. Note that both FM-Mem and W-Mem should be word writable to support writing to a section of a row without changing the value of other memory bits in the same row. The input features from each batch is written to the $\lceil(I/(W_{FM-mem}/B))\rceil= \lceil(200/(64/2))=7\rceil$ rows. At processing time, using the NPE(2,64) configuration, the TCD-NPE in one access (Reading one row) will receive $W_F/B$ input features from $B$ different batches and store them in a buffer. In each subsequent cycle, it consumes one input from each batch, hence, the arrangement of data and sequential read of data into a buffer will reduce the number of memory accesses by a factor of $W_{FM-mem}/B=64/2=32$. \vspace{1mm}

\subsubsection{\textbf{Local Distribution Network (LDN)}}
\begin{figure}[h]
	\centering

	\includegraphics[width=0.8\columnwidth]{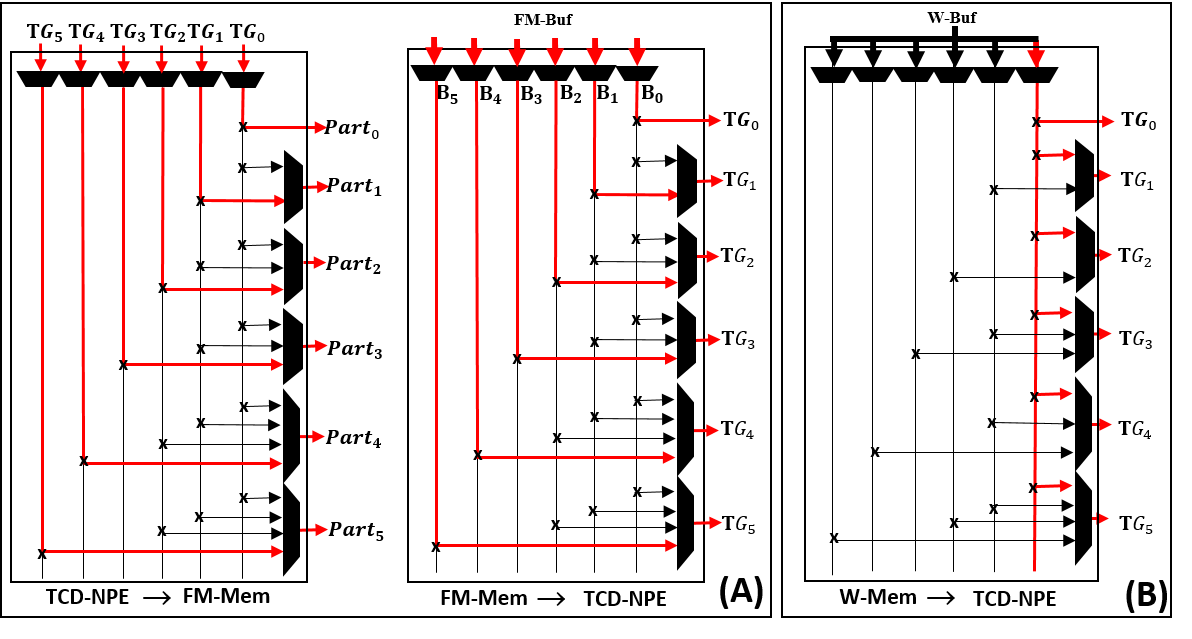}
	\caption{An example of LDN for managing the connection between a ($6 \times 3$)-PE-array's NoC and memory. (A).left: LDN for writing from NoC data bus to FM-mem. (A).right: LDN for reading from FM-mem to NoC bus. (B):  LDN for reading from W-mem into NoC filter bus. The FM-mem in this case, is divided into 6 partitions, supporting the simultaneous process of 6 batches at a time.\vspace{-4mm}}
	\label{ldn}
\end{figure}

The Local Distribution Networks (LDN) interface the read/write buffers and the Network on Chip (NOC). They manage the desired multi- or uni-casting scenarios required for distributing the filter values and feature values across TGs. Figure \ref{ldn} illustrate an example of LDNs in an NPE constructed using $6 \times 3$ array of TCD-MACs. As illustrated in this example, the LDNs are used for 1) reading/writing from/to buffers of FM-mem while supporting the desired multi-/uni-casting configuration (generated by controller) to support the selected NPE(K, N) configuration (Fig.\ref{ldn}.A) and 2) reading from W-mem buffer and multi-/uni-casting the result into TGs (Fig.\ref{ldn}.B). Note that the LDN in Fig, \ref{ldn} is specific to NPE of size $6 \times 3$. For other array sizes, a similar LDN should be constructed.  

\section{Results\vspace{-1mm}}\label{result_section}
In this section, we first evaluate the Power, Performance, and Area (PPA) gain of using TCD-MAC, and then evaluate the impact of using the TCD-MAC in our proposed TCD-NPE. The TCD-MAC and all MACs evaluated in this section operate on signed 16-bit fixed-point inputs.

\subsection{Evaluation and Comparison Framework}\label{framwwork}
The PPA metrics are extracted from the post-layout simulation of each design. Each MAC is designed in VHDL, synthesized using Synopsis Design Compiler \cite{dc} using 32nm standard cell libraries, and is subjected to physical design (targeting max frequency) by using the Synopsys reference flow in IC Compiler \cite{icc}. The area and delay metrics are reported using Synopsys Primetime \cite{pt}. The reported power is the averaged power across 20K cycles of simulation with random input data that is fed to Prime timePX \cite{pt} in FSDB format. The general structure of MACs used for comparison is captured in Fig. \ref{GV_TCD-MAC}. We have compared our solution to a wide array of MACs. In these MACs, for multiplication, we used Booth-Radix-N (BRx2, BRx4, BRx8) and Wallace implementations. For addition we have used Brent-Kung (BK) and Kogge-Stone (KS) adders. Each MAC is identified by the tuple (Multiplier choice, Adder choice).

\begin{table}
	\scriptsize
	\centering
	\caption{PPA comparison between various MACs and TCD-MAC. \vspace{-2mm}}
	\scalebox{0.9}{
		\setlength\tabcolsep{2pt}
		\begin{tabular}{|c|c|c|c|c|}             \hline 
			\textbf{MAC Type} & \textbf{Area}($\mu m^2$ ) & \textbf{Power}($\mu w$) & \textbf{Delay}($ns$) & \textbf{PDP}($pJ$) \tabularnewline            \hline 
			(BRx2, KS)        & 8357                      & 467                     & 2.85                 & 13.31   \tabularnewline                              
			(BRx2, BK)        & 8122                      & 394                     & 3.3                  & 13     \tabularnewline                               
			(BRx8, BK)        & 7281                      & 383                     & 3.14                 & 12.03     \tabularnewline                            
			(BRx4, BK)        & 6437                      & 347                     & 3.35                 & 11.62    \tabularnewline                             
			(WAL, KS)         & 7171                      & 346                     & 3.04                 & 10.52   \tabularnewline                              
			(WAL, BK)         & 6520                      & 334                     & 3.13                 & 10.45     \tabularnewline                            
			(BRx4, KS)        & 6551                      & 393                     & 2.47                 & 9.71   \tabularnewline                               
			(BRx8, KS)        & 7342                      & 354                     & 2.63                 & 9.31   \tabularnewline                               
			TCD-MAC           & 5004                      & 320                     & 1.57                 & 5.02    \tabularnewline                              
			\hline 
		\end{tabular}
	}
	\label{TCD_mac_PPE}
\end{table}

\subsection{TCD-MAC PPA assessment} \label{MAC_PPA}
Table \ref{TCD_mac_PPE} captures the PPA comparison of the TCD-MAC against a popular set of conventional MAC configurations. As reported, the TCD-MAC has a smaller overall area, power and delay compare to all reported MACs. Using TCD-MAC provide 23\% to 40\% reduction in area, 4\% to 31\% improvement in power, and an impressive 46\% to 62\% improvement in PDP when compared to other reported conventional MACs. 

Note that this improvement comes with the limitation that the TCD-MAC takes one extra cycle to generate the correct output when working on a stream of data. However, the power and delay saving of TCD-MAC significantly outweigh the delay and power for one extra computational cycle. To illustrate this, the throughput and energy improvement of using a TCD-MAC for processing different sizes of input streams (1, 10, 100, 1000) is compared against selected conventional MACs and is reported in Table \ref{impCAC_MAC}. As illustrated, when using the TCD-MAC for processing an array of inputs, the power and delay savings quickly outweigh the delay and power of the added cycle as input stream size increases.

\begin{table}[t]
	\scriptsize
	\centering
	\caption{Percentage improvement in throughput and energy when using a TCD-MAC (as opposed to a conventional MAC) to process an stream of 1, 10, 100 and 1000  multiplication and addition operations.  }
	\scalebox{0.9}{
		\begin{tabular}{|c|c|c|c|c|c|c|c|c|}
			\hline 
			\multirow{2}{*}{\textbf{Mac Type}} & \multicolumn{4}{c|}{\textbf{Throughput improvement(\%)}} & \multicolumn{4}{c|}{\textbf{Energy     Improvement(\%)}}\tabularnewline
			\cline{2-9} 
			         & \textbf{1} & \textbf{10} & \textbf{100} & \textbf{1000} & \textbf{1} & \textbf{10} & \textbf{100} & \textbf{1000}\tabularnewline 
			\hline 
			(BRX2, KS) & 25         & 59          & 62           & 63            & -10        & 40          & 45           & 45\tabularnewline            
			(BRX2, BK) & 23         & 58          & 62           & 62            & 5          & 48          & 52           & 53\tabularnewline            
			(BRX8, BK) & 17         & 55          & 58           & 59            & 0          & 45          & 50           & 50\tabularnewline            
			(BRX4, BK) & 14         & 53          & 57           & 57            & 7          & 49          & 53           & 54\tabularnewline            
			(WAL, KS)  & 5          & 48          & 52           & 53            & -3         & 44          & 48           & 49\tabularnewline            
			(WAL, BK)  & 4          & 48          & 52           & 52            & 0          & 45          & 50           & 50\tabularnewline            
			(BRX4, KS) & -3         & 44          & 48           & 49            & -27        & 31          & 36           & 37\tabularnewline            
			(BRX8, KS) & -7         & 41          & 46           & 47            & -19        & 35          & 40           & 41\tabularnewline            
			\hline 
		\end{tabular}
	}
	
	\label{impCAC_MAC}
\end{table}

\subsection{TCD-NPE Evaluation}
In this section, we describe the result of our TCD-NPE implementation as described in section \ref{TCD_NPE_section}. Table \ref{TCD_NPE_spec}-top summarizes the characteristics of TCD-NPE implemented, the result of which is reported and discussed in this section. For physical implementation, we have divided the TCD-NPE into two voltage domains, one for memories, and one for the PE array. This allows us to scale down the voltage of memories as they had considerably shorter cycle time compared to that of PE elements. This choice also reduced the energy consumption of memories and highlighted the saving resulted from the choice of MAC in the PE-array. Note that the scaling of the memory voltage could be even more aggressive than what implemented in our solution; In several prior work \cite{sasan2,sasan3,sasan4,sasan5,sasan6}, it was shown that it is possible to significantly reduce the read/write/retention power consumption of a memory unit by aggressively scaling it supplied voltage while deploying architectural fault tolerance techniques and solutions to mitigate the increase in the memory write/read/retention failure rate. On top of that, learning solutions are also approximate in nature, and inherently less sensitive to small disturbance to their input features. This inherent resiliency could be used to deploy fault tolerant techniques to only protect against bit errors in most significant bits of input feature map, resulting in reduced complexity of deployed fault tolerance scheme. 

Table \ref{TCD_NPE_spec}-bottom captures the overall PPA of the implemented TCD-NPE extracted from our post layout simulation results which are reported for a Typical Process, at 85C$^{\circ}$ temperature, when the PE-array and memory elements voltages are set according to Table \ref{TCD_NPE_spec}.

\begin{table}[t]
	\scriptsize
	\centering
	\caption{TCD-NPE implementation details and PPA results. In this table, we have only reported the leakage power. The dynamic power is activity dependent. The breakdown of energy consumption for processing different benchmarks is reported in Fig. \ref{benchmarking}}
	\scalebox{0.9}{
		\setlength\tabcolsep{2pt}
		\begin{tabular}{|l|l|l|l|}             \hline 
			\textbf{Feature}    & \textbf{Detail}                            & \textbf{Feature}            & \textbf{Detail}  \tabularnewline            \hline 
			\textbf{PE-array}   & $16 \times 8$             & \textbf{Processing Element} & TCD-MAC         \tabularnewline                    
			\textbf{Input Data Format} & Signed 16-bit fixed-point               &  \textbf{Data Flow}  & OS                  \tabularnewline   
			\textbf{W-mem size} & $512$ KByte                                   & \textbf{Activation Units}   & Relu               \tabularnewline                 
			 \textbf{FM-mem Size}        & $2 \times 64$ KByte & \textbf{PE-array voltage}   & $0.95$V              \tabularnewline                 
			\textbf{Mapper}     & Off-chip using Alg. \ref{mapper_algorithm}                                          & \textbf{Mem voltage}        & $0.70$V                 \tabularnewline                 \hline
			\textbf{Area}                   & $3.54$ mm$^2$     & \textbf{Max Frequency}        & $636$ MHz          \tabularnewline                   
			\textbf{PE-array Area}          & $0.724$ mm$^2$    & \textbf{Memory Area}          & $2.5$ mm$^2$           \tabularnewline              
			\textbf{Overall Leak. Power}  & $75.5$ mW          & \textbf{Memory Leak. Power} & $51.7$ mW                    \tabularnewline          
			\textbf{PE-array Leak. Power} & $6.4$ mW           & \textbf{Others Leak. Power} & $17$ mW              \tabularnewline                 
			\hline 
		\end{tabular}
	}
	\label{TCD_NPE_spec}
    \vspace{-2 mm}
\end{table}

			

\begin{figure}
	\centering
	\includegraphics[width=0.80\columnwidth]{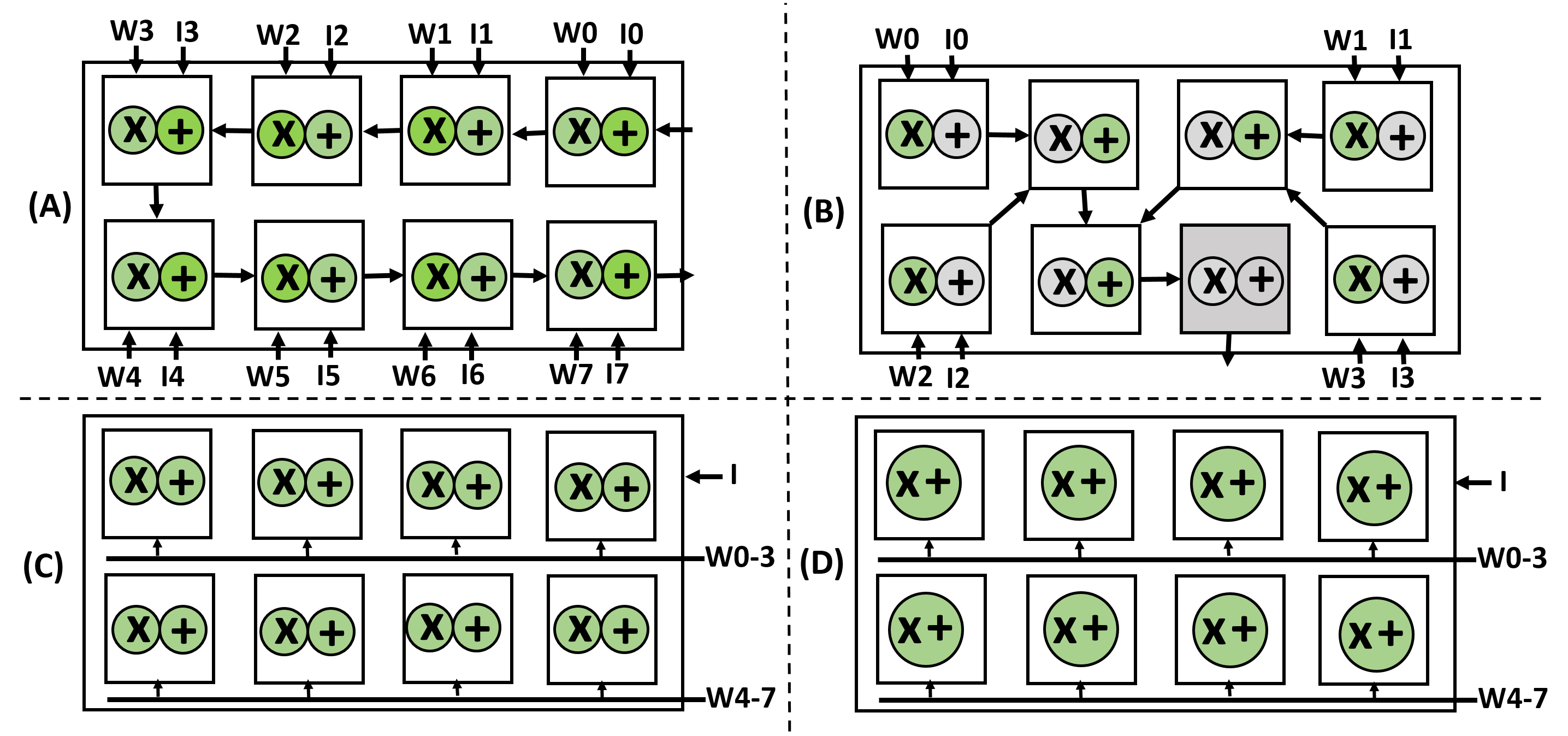}
	\caption{Four possible data flow for processing an MLP model. (A): NLR data flow using conventional MACs to form a systolic array. (B): RNA data flow resulted from unrolling the MLP model and mapping the computation tree to conventional MACs (each used as either multiplier or adder) as described in \cite{tu2015rna}. (C) The OS data flow using conventional MAC. (D): The OS dataflow using TCD-MAC.}
	\label{data_flow_comp}
\end{figure}

To compare the effectiveness of TCD-NPE, we compared its performance with a similar NPE which is composed of conventional MACS. According to the discussion in section \ref{related_work}, we limit our evaluation to the processing of MLP models. Hence, the only viable data flows are OS and NLR. The TCD-MAC only supports OS, however, by replacing a TCD-MAC with a conventional MAC, we can also compare our solution against OS and NLR. We compare 4 possible data flows that are illustrated in Fig. \ref{data_flow_comp}.  In this Fig. The case (A) is NLR data flow (supported only by conventional MAC) for computing the Neuron values by forming a systolic array withing the PE-array. The case (B) An NLR data flow variant according to \cite{tu2015rna} when the computation tree is unrolled and mapped to the PEs, forcing the PE to either act as an adder or multiplier. The case (C) is the OS data flow realized by using conventional MAC. And, finally, the case (D) is the OS data flow implemented using TCD-NPE.  

For OS dataflows, we have used the algorithm \ref{mapper_algorithm} to schedule the sequence of computational rounds. We have compared the efficiency of each of four data flows (described in Fig. \ref{data_flow_comp}) on a selection of popular MLP benchmarks characteristic of which is described in Table. \ref{benchmark}.

\begin{table}[h]
	\scriptsize
	\centering
	\caption{MLP benchmarks used in this work \cite{Dua:2019}.  }
	\scalebox{0.9}{
		\begin{tabular}{|l|l|l|}
			\hline 
			\textbf{Applications} & \textbf{Dataset} & \textbf{Topology}\tabularnewline  
			\hline 
			Digit Recognition     & MNIST            & 784:700:10\tabularnewline         
			Census Data Analysis  & Adult            & 14:48:2\tabularnewline            
			FFT                   & Mibench data     & 8:140:2\tabularnewline            
			Data Analysis         & Wine             & 13:10:3\tabularnewline            
			object Classification & Iris             & 4:10:5:3\tabularnewline           
			Classification        & poker Hands      & 10:85:50:10\tabularnewline        
			Classification        & Fashion MNIST    & 728:256:128:100:10\tabularnewline 
			\hline
		\end{tabular}
	}
	\label{benchmark}
\end{table}

\begin{figure}[!hbt]
	\centering
	\includegraphics[width=0.98\columnwidth]{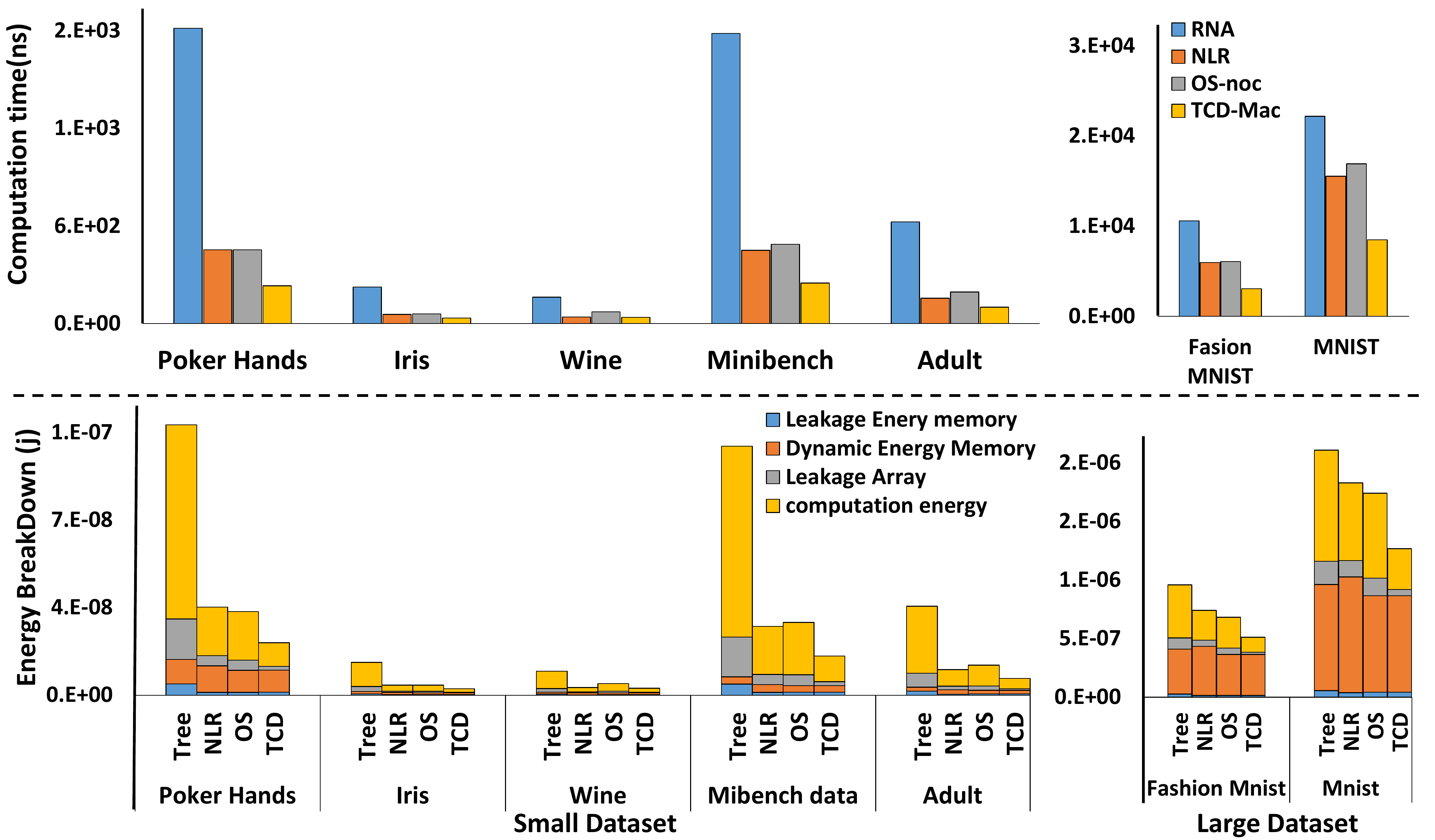}
	\caption{Comparison of TCD-NPE with an NPE constructed using conventional MACs that uses the OS, NLR, or RNA data flow. top): Execution time for various MLP benchmarks. Bottom): Energy consumption for various MLP benchmarks.\vspace{-4mm}}
	\label{benchmarking}
	   
\end{figure}

As illustrated in Fig. \ref{benchmarking}.left, the execution time of the TCD-NPE is almost half of an NPE that uses a conventional MAC in either OS or NLR data flow, and significantly smaller than the RNA data flow (an NLR variant) that was proposed in \cite{tu2015rna}. Fig. \ref{benchmark}.right captures the energy consumption of the TCD-NPE and compares that with a similar NPE constructed using conventional MACs. For each benchmark, the energy consumption is broken into 1) computation energy of PE-array, 2) the leakage of the PE-array, 3) the leakage of the memory, and 4) the dynamic energy of memory (and buffer combined). Note that the voltage of the memory is scaled to a lower voltage, as described in table \ref{TCD_NPE_spec}. This choice was made as the cycle time of the PE's was significantly shorter than the memory cycle times. The scaling of the memory voltage increased its associated cycle time to one cycle, however, significantly reduced its dynamic and leakage power, making the PE-array energy consumption the largest energy consumer. In addition, note that by sequentially shaping the data in the memories, and usage of buffers, we significantly reduced the number of required memory accesses, resulting in a significant reduction in the dynamic power consumption of the memories. As illustrated, the TCD-NPE not only produces the fastest solution but also produces the least energy-consuming solutions across all NPE configurations, all data flows and all simulated benchmarks. 

\section{Conclusion}

In this paper, we introduced the concept of temporal carry bits and used the concept to design a novel MAC for efficient stream processing (TCD-MAC). We further proposed the design of a Neural Processing Engine (TCD-NPE) that is architected using an array of TCD-MACs as its processing element. We, further, proposed a novel scheduler that schedules the sequence of events to process an MLP model in the least number of computational rounds in the proposed TCD-NPE. We reported that the TCD-NPE significantly outperform similar neural processing solutions that are constructed using conventional MACs in terms of both energy consumption and execution time (performance). \vspace{-2 mm}

\nocite{icc}
\nocite{dc}
\nocite{pt}
\nocite{rh}

\renewcommand{\IEEEbibitemsep}{0pt plus 0.5pt}
\makeatletter
\IEEEtriggercmd{\reset@font\normalfont\fontsize{8.1pt}{8.1pt}\selectfont}
\makeatother
\IEEEtriggeratref{1}

\bibliographystyle{IEEEtran}
\bibliography{main}


\end{document}